\documentclass[letter]{aa}  

\usepackage{graphicx}
\usepackage{txfonts}

\usepackage{color, colortbl}

\usepackage{amstext}

\begin{document} 

\definecolor{bubbles}{rgb}{0.91, 1.0, 1.0}
\definecolor{columbiablue}{rgb}{0.61, 0.87, 1.0}
\definecolor{cream}{rgb}{1.0, 0.99, 0.82}
\definecolor{lightblue}{rgb}{0.68, 0.85, 0.9}
\definecolor{lightcyan}{rgb}{0.88, 1.0, 1.0}

   \title{Bright single-mode RR Lyrae stars: matching Gaia EDR3 with 
          pulsation and evolutionary models}

   \author{Geza Kovacs
          \inst{}
          and 
          Behrooz Karamiqucham
          \inst{}
          }

   \institute{Konkoly Observatory,  Research Center for Astronomy and Earth Sciences, E\"otv\"os Lor\'and Research Network \\ Budapest, 1121 Konkoly Thege ut. 15-17, Hungary\\
              \email{kovacs@konkoly.hu}
             }

   \date{Received September 09, 2021; accepted ...}


%
%
  \abstract
{We combine observed metallicity, optical and infrared magnitudes 
with evolutionary and pulsation models to derive average luminosities 
for $156$ single-mode RR~Lyrae stars. These luminosities are compared 
with those obtained from the Gaia EDR3 parallaxes, and found in 
excellent agreement with the high accuracy subsample ($62$ stars, 
with relative parallax errors less than $2$\%). With the temperature 
and metallicity scale used, no parallax shift seems to be necessary 
when $\alpha$-enhanced evolutionary models are employed. Some $10$\% 
of the sample shows curious `distance keeping' between the evolutionary 
and pulsation models. The cause of this behavior is not clear at this 
moment but can be cured by an excessive increase of the reddening.}

   \keywords{Stars: fundamental parameters -- 
             stars: distances -- 
             stars: variables: RR~Lyrae -- 
             stars: oscillations --
             stars: horizontal branch
               }

\titlerunning{Single-mode RR Lyrae stars and Gaia EDR3}
\authorrunning{Kovacs, G. \& Karamiqucham, B.}

   \maketitle
%
%
%
\section{Introduction}
\label{sect:intro}
In a former paper \citep[][hereafter K21]{kovacs2021} we used 
pulsational and evolutionary models to close the relations 
governing the basic physical parameters of double-mode RR~Lyrae 
(RRd) stars. The so-derived luminosities of $30$ galactic RRd 
stars were compared with the direct estimates from the Gaia 
Early Data Release 3 \citep[EDR3,][]{lindegren2021}. The agreement 
between these independent sets was satisfactory, considering 
the overall low brightness of the available objects and the 
consequentially low accuracy of their parallaxes. 

One of the issues in the case of RRd stars is the lack of 
chemical composition measurements. In K21 we determined the overall 
metal abundance [M/H] as a by-product of the matching procedure 
of the pulsation and the evolutionary models. Opposite to the RRd 
stars, single-mode RR~Lyrae (RRabc) stars have very abundant chemical 
analyses, many of them are based on  high-dispersion spectra and 
multiple visits by different studies. The work presented in this 
paper heavily relies on past and current spectral analyses by fixing 
[M/H] and thereby letting us to use RRabc stars in matching the 
pulsation and evolutionary models.  

The ultimate goal of this work is to check the consistency of 
the nearly directly derived Gaia luminosities with the model 
values by using basic observables only. Because of the 
significantly higher accuracy of the input data and the 
weaker dependence on the more intricate details of the pulsation 
models, we consider this study to be more stringent for the 
compatibility of the Gaia and theoretical luminosities than the 
one presented in K21 (even though the two studies are consonant 
within the error limits).

%
%
\section{Method}
\label{sect:method}
\begin{figure}[h]
\centering
\includegraphics[width=0.35\textwidth]{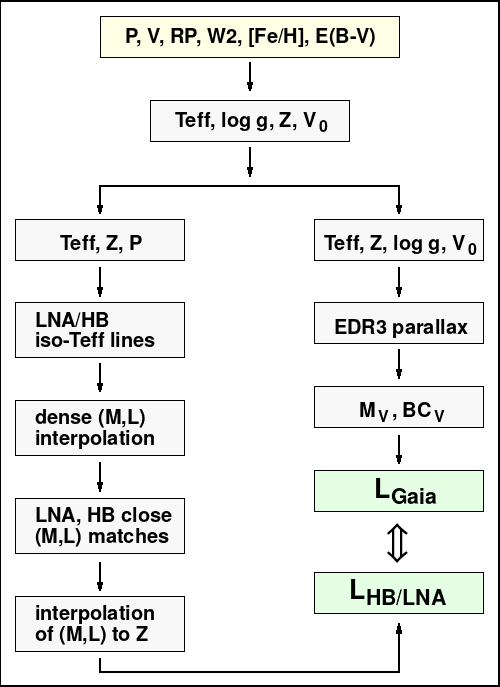}
\caption{Main steps of the star-by-star procedure to obtain 
         estimates of the RR~Lyrae luminosities. Boxes on the 
	 left exhibit the steps to compute the theoretical 
	 luminosity $\rm L_{\rm HB/LNA}$, whereas those on the 
	 right show the steps leading to the empirical 
	 luminosity $\rm L_{\rm Gaia}$. We follow standard 
	 notation (i.e., $RP$ for the Gaia red and $W2$ for 
	 the WISE W2 magnitudes).} 
\label{rr1_chart}
\end{figure}

Our method of deriving the basic physical parameters (mass, luminosity, 
temperature -- $M$, $L$, $T_{\rm eff}$, respectively) is similar to 
those used in our earlier papers \citep[][K21]{kovacs1999, kovacs2000}. 
A simple flowchart of the analysis is shown in Fig.~\ref{rr1_chart}. 
Some additional details are as follows. 
%
%
%

For any given star, we have the following input parameters: 
period $P$, intensity-averaged Johnson V and K magnitudes and 
iron abundance [Fe/H] (see Sect.~\ref{sect:data} for the derivation 
of these quantities). From these parameters we calculate 
$T_{\rm eff}$ -- the weak gravity ($\log g$) dependence can be 
handled via the pulsation equation as given in K21. We use the 
$T_{\rm eff}$ zero point of \cite{gonzalez2009} as detailed in 
K21.  

Once $T_{\rm eff}$ has been estimated, for its fixed value we can 
compute the corresponding $(M,L)$ values for the evolutionary and 
pulsation models. To avoid the somewhat cumbersome procedure of 
directly interpolating the model values to the observed metallicity, 
we opted to compute the iso-$T_{\rm eff}$ lines for three metallicity 
values bracketing the observed metallicity. The following metric 
is used to measure the distance between the pulsation and evolution 
models for any matching metallicity $Z$ 
%
%
%
\begin{eqnarray}
\label{eq_ml_match}
{\cal D}(M,L,Z) = \sqrt{\log^2 \Biggl({L^{HB} \over L^{LNA}}\Biggr) + \log^2 \Biggl({M^{HB} \over M^{LNA}}\Biggr)} \hspace{2mm} ,
\end{eqnarray}
where the superscripts $HB$ and $LNA$ denote, respectively, the 
horizontal branch evolution and linear non-adiabatic pulsation 
models. Once the minima of ${\cal D}(M,L,Z)$ for the three 
metallicities are found, we use quadratic interpolation in $\log(Z)$ 
to derive the closest matching $(M,L)$ values for the $HB$ and $LNA$ 
models. 

We use the recently updated $HB$ models from the Bag of Stellar Tracks 
and Isochrones (BaSTI) project, as described by \cite{hidalgo2018}. 
The models have fixed metallicity values spanning the full range of 
galactic field RR~Lyrae stars (i.e., $-3.0<$~[Fe/H]~$<0.1$). All models 
are without convective overshooting. We use two sets of models: those 
with scaled solar-composition and those with enhanced $\alpha$-elements 
(i.e., O, Ca, Mg, etc). In the latter case, the degree of enhancement 
is [$\alpha$/Fe]$=+0.4$ and all models have diffusion and mass loss 
of $\eta=0.3$ (none of the solar-scaled models we use have either of 
these). In searching for the best HB/LNA matches, we consider all 
evolutionary stages, not only those close to the zero-age HB.   

The pulsation models are linear non-adiabatic models, without 
$\alpha$-enhancement -- the effect of enhancement on the period is in 
the range of few times $10^{-3}$~days, too small to have an influence 
on the derived stellar parameters. All models have a Hydrogen abundance 
of $0.76$, very close to those of the $HB$ models, and matching $Z$ 
values to the $HB$ metallicity grid (altogether $18$ discrete metallicities). 
For any fixed $Z$, there are $4095$ models on the $(M, L, T_{\rm eff})$ 
grid given in Table~2 of K21. 

Since we use observed [Fe/H], we need to convert this to the corresponding 
overall metallicity $Z$ for the enhanced models. We do this conversion by 
using the table available at the BaSTI home page. The $+0.4$~dex increase 
in the $\alpha$-elements results in a factor of two overall increase in 
$Z$ throughout the full range of [Fe/H].

%
%
\section{Datasets}
\label{sect:data}
Most of the RR~Lyrae stars we deal with in this paper are bright,  
($8 < V < 13$) and therefore, have abundant observational 
material. Nevertheless, various observational settings and details 
in the analyses may introduce differences in the derived parameters. 
Consequently, in gathering the essential data we tried to rely on 
homogeneous datasets. For [Fe/H] they are apparently not available, 
therefore, we rescaled the data from the same source and adjusted 
the zero point to some chosen level. 

Our basic dataset contains $67$ stars with EDR3 parallax errors less 
than $2$\%. For assessing error dependence we also use a larger 
dataset, containing additional $94$ stars with larger relative 
parallax errors, up to $10$\%. Both sets are based on the combination 
of the target lists of \cite{dambis2013}, \cite{monson2017} and 
several recent spectroscopic studies, detailed in Sect.~\ref{sect:feh}. 
The basic set is complete in respect of the above source lists 
with the parallax constraint. The extended set is incomplete, and 
satisfies mostly the constraints of ``relatively low parallax error'', 
``easy data access'' and [Fe/H] availability. 

%
%
\subsection{Iron abundance}
\label{sect:feh}
In recent years, several high-dispersion spectroscopic (HDS) works 
have been devoted to deep chemical analysis of RR~Lyrae stars. Here 
we use the combined lists of \cite{magurno2018} and \cite{crestani2021}, 
based also on the works of 
\cite{clementini1995}, \cite{fernley1996}, \cite{lambert1996}, 
\cite{for2011}, \cite{hansen2011}, \cite{liu2013}, 
\cite{nemec2013}, \cite{govea2014}, \cite{pancino2015}, 
\cite{chadid2017}, \cite{sneden2017} and \cite{andrievsky2018}.  
The individual [Fe/H] values in these compilations may differ 
by as much as $\sim 0.4$~dex, due to, e.g., differences in the 
reduction methods or the turbulent velocity used. Therefore, 
simple averaging is, in general, does not seem to be justified. 
Consequently, we followed the procedure below to put all [Fe/H] 
on the same scale. 

%
%
\begin{figure}[h]
\centering
\includegraphics[width=0.45\textwidth]{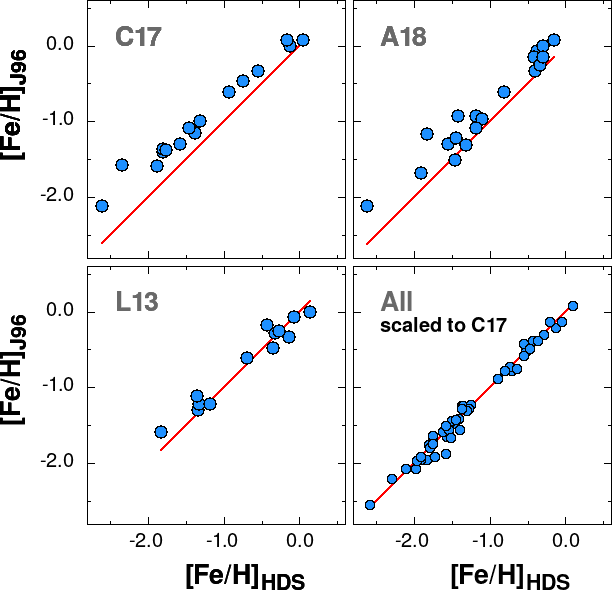}
\caption{Published iron abundances: \cite{jurcsik1996} vs 
         high-dispersion spectroscopic [Fe/H] for the data 
	 sources given in the upper left corner of each panel 
	 \citep[C17, A18 and L13 for][respectively]{chadid2017, andrievsky2018, liu2013}. 
	 In the panel at the lower right we plot the transformed 
	 values for all [Fe/H] matching the list of 
	 \cite{jurcsik1996}. The HDS identity line is shown 
	 by red continuous line.} 
\label{jk96_hds}
\end{figure}

To establish a common base of comparison, we used the metallicities 
compiled by \cite{jurcsik1996} (hereafter J96). This list is 
based on the low-dispersion spectroscopic (LDS) measurements 
primarily by \cite{blanco1992}, \cite{layden1994} and 
\cite{suntzeff1994}. We found that the J96 metallicities correlate  
well with the values of the individual HDS studies but there are 
systematic differences in the zero points (ZPs) and that the HDS 
values are lower toward lower [Fe/H]. In particular, for 
\cite{chadid2017} (hereafter C17) we find that 
%
%
%
\begin{eqnarray}
\label{j96}
{\rm [Fe/H]}_{\rm C17} = 1.10{\rm [Fe/H]}_{\rm J96} - 0.22 \hspace{2mm} ,
\end{eqnarray}
with an RMS scatter of $0.12$~dex.\footnote{Omitting the single 
outlier SS~Leo, we have $\sigma=0.07$~dex.} This excellent agreement 
(see Fig.~\ref{jk96_hds}) between the two independent sets of 
abundances led us to transform all HDS and the LDS datasets of 
\cite{layden1994} and \cite{layden1996} to the scale of C17, and 
then, average them out for each target. For the HDS sets we employed 
only ZP shifts. Several datasets did not have overlap with C17. 
Then, we used the overlaps with those that had overlaps with the 
set in question and with C17 to transform the published [Fe/H] 
for that dataset. Altogether we ended up with $177$ objects, most 
of them with HDS metallicities, often with independent 
measurements from multiple (up to six) sources per target.

%
%
\subsection{Reddening}
\label{sect:reddening}
One of the sensitive parameters entering in the estimation of the 
luminosity is interstellar reddening. Because the commonly used 
all-sky maps give the total reddening at a certain direction, the 
exact amount of the extinction between the object and the observer 
remains hidden. In the case of excessive reddening this might lead 
to a serious overestimation of the reddening. For instance, based 
on the map of \cite{schlafly2011}, we got over $10^4$~K for 
$T_{\rm eff}$ in the case of RZ~Cep, AR~Per, BN~Vul, SW~Cru and BI~Cen 
with $E(B-V)>0.7$ for all of them. 

For RRab stars one may use some of the formulae available to estimate 
their intrinsic colors \citep{sturch1966, blanco1992, kovacs1997}. 
We found this approach unsatisfactory, mainly because of the need 
of additional color and light curve information and the relative 
low accuracy of the derived reddenings. The ideal approach would be 
to use spectroscopic $T_{\rm eff}$ and employ the related intrinsic 
color index to estimate $E(B-V)$ from the observed colors. Although 
HDS [Fe/H] values are often derived from multiple epoch data, the 
phase coverage is usually rather limited, and it is out of the scope 
of this paper to derive reliable average $T_{\rm eff}$ values from 
those sparsely-sampled sets. 

We also tested the recent 3D map of \cite{green2019} derived from 
2MASS and Pan-STARRS 1 colors, aided by the Gaia parallaxes. 
Except for the targets with large reddenings, we found no improvement 
in the correlation between the HB/LNA and Gaia luminosities. 
Consequently, we stay with the map of \cite{schlafly2011} accessible 
at the NASA/IPAC Infrared Science 
Archive.\footnote{\url{https://irsa.ipac.caltech.edu/applications/DUST/}}

%
%
\subsection{Johnson V, K photometry}
\label{sect: VK}
The intensity-averaged $V$ magnitudes have been gathered from three 
sources. In the order of priority, they come from \cite{monson2017},  
\cite{dambis2013} and from ASAS \citep{pojmanski1997}, whenever the 
given target was not available in the first two 
sources.\footnote{For the ASAS light curves we fitted Fourier series 
of order $4$--$10$ to derive the intensity-averaged $V$ magnitudes 
-- see K21 for further details.}  

For the intensity-averaged 2MASS $K$ ($Ks$) magnitudes we could use 
the recently compiled set of \cite{layden2019}. However, even though 
it is the biggest homogenized sample of direct $Ks$ measurements 
available today, it misses some Gaia objects with high-precision 
parallaxes (e.g., CS~Eri, EL~Aps, etc). In addition, several 
$Ks$ values have been derived via template fitting due to the low 
number of data points available. As a result, we inclined to use 
the unWISE catalog of \cite{schlafly2019} and sought for a 
transformation from the WISE and Gaia simple average magnitudes to 
approximate the $Ks$ values of \cite{layden2019}. With the $145$ 
stars used, we found that by leaving out eight $2.5\sigma$ outliers, 
the following formula fits Layden's et al. data with an RMS scatter 
of $0.028$~mag
%
%
%
\begin{eqnarray}
\label{K_fit}
Ks & = & \phantom{(}9.958\pm0.004   \nonumber \\ 
   & + & (0.884\pm0.028)(W2-9.949)  \nonumber \\
   & + & (0.120\pm0.030)(RP-10.736) \hspace{2mm} .
\end{eqnarray}
Here $RP$ denotes the Gaia EDR3 red magnitude. The unWISE $W2$ 
magnitude is calculated from the published FW2 fluxes from 
$W2=22.5 - 2.5\log_{10}(FW2)$. Unlike the transformation presented 
in K21, Eq.~\ref{K_fit} is applicable also to RRab stars, due to 
the correction introduced by the Gaia red magnitude.  
 
%
%
\section{Luminosities}
\label{sect:lumin}
\begin{figure}[h]
\centering
\includegraphics[width=0.45\textwidth]{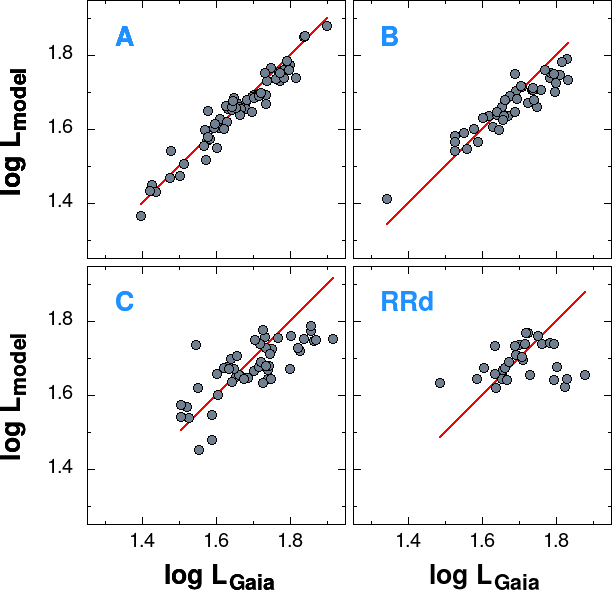}
\caption{Gaia EDR3 vs evolutionary and pulsation model luminosities for 
         single-mode RR~Lyrae stars (panels A, B, C). The single-letter 
	 labels denote independent datasets in the order of increasing 
	 parallax error (see Table~\ref{LL_tab}). We used $\alpha$-enhanced 
	 evolutionary models. For comparison, we also plotted the 
	 parallax-shifted RRd models of K21. The Gaia identity lines are 
	 shown by red continuous lines.} 
\label{LL_pl}
\end{figure}

To investigate the data quality dependence of the level of agreement 
between the Gaia and model (HB/LNA) luminosities, we divided the total 
sample into three groups. The division is based on the relative parallax 
errors (see Table~\ref{LL_tab}). Set A (the basic dataset without 
the five stars with overestimated reddenings -- see Sect~\ref{sect:reddening}) 
is the most populated and of the highest accuracy. Figure~\ref{LL_pl} 
shows the correlation between the Gaia and model luminosities following 
the division mentioned and completed by a similar plot for the RRd 
stars from K21. 
%
%
\begin{table}[h]
\centering
\begin{minipage}{200mm}
\caption{Gaia and model luminosities}
\label{LL_tab}
\scalebox{1.0}{
\begin{tabular}{lcccc}
\hline
\multicolumn{5}{c}{$\alpha$-enhanced models} \\
\hline
Set    & $(r_{\rm min},r_{\rm max})$      & $\langle \Delta \log L \rangle$ & RMS & N \\
\hline 
A      & $(0.00,0.02)$ & $\phantom{-} 0.006 \pm 0.004$ & 0.029 & 62 \\
B      & $(0.02,0.03)$ & $\phantom{-} 0.015 \pm 0.006$ & 0.041 & 43 \\ 
C      & $(0.03,0.10)$ & $\phantom{-} 0.026 \pm 0.009$ & 0.062 & 51 \\
\hline
\multicolumn{5}{c}{solar-scaled models} \\
\hline
A0     & $(0.00,0.02)$ & $-0.012 \pm 0.004$               & 0.030 & 62 \\
B0     & $(0.02,0.03)$ & $-0.003 \pm 0.006$               & 0.042 & 43 \\
C0     & $(0.03,0.10)$ & $\phantom{-} 0.007 \pm 0.009$    & 0.063 & 51 \\
\hline
\multicolumn{5}{c}{solar-scaled models} \\
\hline
RRd    & $(0.03,0.13)$ & $\phantom{-} 0.015 \pm 0.016$    & 0.087 & 30 \\
\hline
\end{tabular}}
\end{minipage}
\begin{flushleft}
\vspace{-5pt}
{\bf Notes:}
$(r_{\rm min},r_{\rm max})$: range of the relative parallax errors --  
$\langle \Delta \log L \rangle$: average of $\log (L_{\rm Gaia}/L_{\rm model})$ --  
$RMS$: standard deviation of the residuals --  
$N$: number of data points in the given set.  
\end{flushleft}
\end{table}
%

%
%

Let us recall that (except for the RRd stars) no adjustments were made 
for any of the parameters used. The dependence on the data quality is 
clearly visible. However, the Gaia luminosities seem to be systematically 
overestimated for higher luminosities. The difference depends on the 
quality of the data and it increases for poorer data. We suspect that 
most of the effect comes from the increasing errors of the Gaia 
luminosities that are, on the average, twice as large as the HB/LNA 
errors. This difference leads to a flattening of the $\log L-\log L$ 
relation, with the Gaia luminosities plotted on the abscissa (see, in 
particular, dataset B in Fig.~\ref{LL_pl}). Then, the finite sample 
size and non-uniform sampling lead to further distortion of the ridge, 
causing apparent systematic deviations for certain groups of stars.  

For set A with accurate parallaxes, the $\alpha$-enhanced models show 
a preference against the solar-scaled models with respect to the 
Gaia luminosities. For sets B and C this preference diminishes, primarily 
because of the higher parallax errors. For set A no parallax shift (with a 
sufficient significance) is needed. In K21 we found a shift of $+0.02$~mas 
to be appropriate for the RRd stars with low overall parallax accuracy. 
However, the significance of this shift is only a few sigma, even 
if we omit the stars with apparent outlier status. The $\alpha$-enhanced 
models perform also better in terms of the HB/LNA solution 
(see Sect.~\ref{sect:dk}), lending further support for to use of these 
models.

%
%
\section{The distance keepers}
\label{sect:dk}
Our model luminosities are based on the solutions provided by 
the intersections of the iso-$T_{\rm eff}$ HB and LNA lines. 
Unfortunately, as in K21, this solution may not exist for some 
observed parameter sets, even if we consider the errors associated 
with the given objects. We refer to these `stubborn' objects as 
``distance keepers'', or DKs.

%
%
\begin{figure}[h]
\centering
\includegraphics[width=0.45\textwidth]{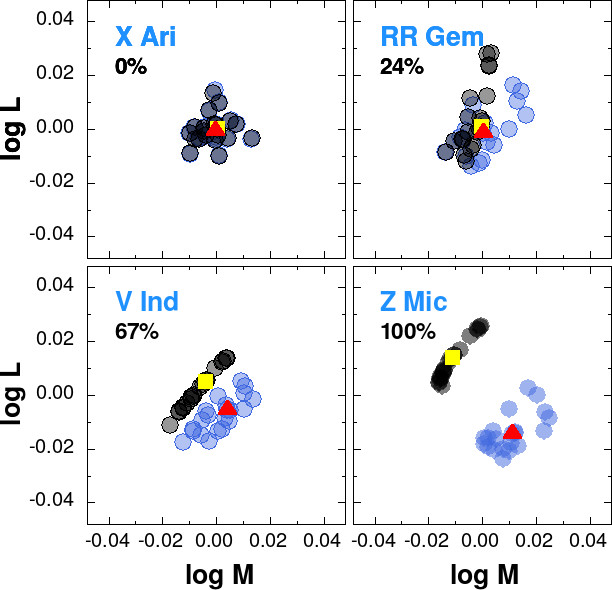}
\caption{Examples on the solution types from the HB and LNA models. 
         Filled circles show the closest $(M,L)$ pairs of the two 
	 types of models from $20$ random realizations with the errors 
	 of $V$, $Ks$, [Fe/H], $E(B-V)$ and parallax, centered around 
	 their observed values. Gray and blue coloring are used 
	 for the HB and LNA models, respectively. The closest values 
	 at the observed parameters are shown by yellow squares 
	 for the HB and by red triangles for the LNA models. 
	 The odds for being a DK are shown under the variable names.} 
\label{sd_plot}
\end{figure}

As a part of the error estimate of the model luminosities, we calculated 
the solutions for each object by using the observed $V$, $Ks$, [Fe/H], 
$E(B-V)$ and parallax values and for $20$ perturbed sets, obtained by 
adding perturbations computed from the observational errors and assuming 
Gaussian distribution for these errors. We searched for the closest 
$(M,L)$ pairs from the HB and LNA models and derived the metric given 
by Eq.~\ref{eq_ml_match} for each realization. These simulations 
yielded not only the errors of the theoretical luminosities but also 
informed us if the high ${\cal D}(M,L,Z)$ values could be explained 
by observational errors. Figure~\ref{sd_plot} shows the various scenarii 
found in our sample. The two upper panels show examples when solutions 
were found for the observed parameters -- albeit with slightly different 
statistical significance. The lower two panels are examples for the DKs. 
Indeed, both stars, although at various degrees, ``keep the distance'' 
between the nearest HB and LNA points in the $(M,L)$ space. 

By (somewhat arbitrarily) fixing the maximum ${\cal D}(M,L,Z)$ to $0.01$,   
we counted the number of realizations for each object exceeding this limit. 
The relative number of these cases to the $20$ realizations ($P_{\rm DK}$) 
for each object yielded an approximation of the probability that a given 
object is a DK. We found that for the $\alpha$-enhanced models $122$ have 
lower than $50$\% chance of being a DK (and, $\sim 100$ have $0$\%). 
However, there are some $15$ stars without HB/LNA solution (i.e., 
practically none of the simulated values yield lower than $0.01$ for 
${\cal D}(M,L,Z)$).  

The analysis of $P_{\rm DK}$ is also useful in the comparison of the 
overall performance of $\alpha$-enhanced models with the solar-scaled 
models. We checked set A (see Table~\ref{LL_tab}) and found that, 
on the average, $P_{\rm DK}$ is smaller for the enhanced models by 
$0.2$ for those with $P_{\rm DK}>0.5$ for the solar-scaled models. 
The number of severe DKs (i.e., those with $P_{\rm DK}>0.9$) shrank 
to $3$ from 7 when switching from the solar-scaled to the enhanced 
models. Some stars with low ${\cal D}(M,L,Z)$ of $0.01-0.02$ have 
completely recovered from the DK status by using enhanced models 
(e.g., SW~And, AV~Peg, U~Pic). We consider these figures as additional 
pieces of evidence for the likely higher relevance of the 
$\alpha$-enhanced models. 

We made a few tests to investigate the possibility of major parameter 
errors causing the DK classification of the more stubborn stars. We 
introduced large and separate changes in the period ($\pm 0.01$), 
$V$ magnitude ($\pm 0.03$), [Fe/H] ($\pm 0.4$) and in $E(B-V)$ (as much 
as it was needed to go below the cutoff for ${\cal D}(M,L,Z)$). We found 
that even with the above large changes in $P$, $V$ and [Fe/H], 
the more stubborn DKs are left in the same status. However, varying 
$E(B-V)$ may help. We examined the top $21$ DKs with $P_{\rm DK}>0.8$. 
We found solutions for each of them by {\em increasing} $E(B-V)$. The 
amount of increase varied between $0.01$ and $0.04$ (well above the 
nominal errors of their reddenings). Two variables (ST~Oph and TV~Lib) 
required even more excessive increase ($0.07$ and $0.10$, respectively). 
Together with its short period and light curve shape, in the case of 
TV~Lib this might indicate a different variable classification (e.g., 
evolved high-amplitude $\delta$~Scuti status). 

Although the all out increase of $E(B-V)$ might indicate the need for 
shifting the zero point of the $T_{\rm eff}$ scale, we found that an 
increase of $0.0045$ in $\log T_{\rm eff}$\footnote{That is, we use 
the $T_{\rm eff}$ scale compatible with the \cite{castelli1997} models 
rather than the recent scale of \cite{gonzalez2009}, based on the 
infrared flux method.} is insufficient to cure the notorious DKs. 
The increased temperature leads to more luminous theoretical models 
and thereby a difference of $-0.015$ in $\log L$ (Gaia minus models). 
Although this can be remedied for the enhanced models of set A by 
an overall parallax decrease of $0.02$~mas, the final gain of these 
zero point shifts does not seem to be justified in seeking the root 
cause of the DK phenomenon. 

One may also wonder if a slight change in the Helium abundance 
(Y) may alter the DK status. Although a deeper discussion of the 
effect of Helium is out of the scope of this study, we note the 
following. Increasing Y increases the HB luminosity 
\citep[e.g.,][]{sweigart1976,marconi2018} and with the pulsation 
periods remaining basically the same, the DK status exacerbates. 
Decreasing Y by $0.02$--$0.05$ may match the evolutionary and 
pulsational luminosities, however, the smaller Helium abundance may 
raise other questions concerning its currently accepted primordial 
value\footnote{Which is Y$\sim 0.245$, used also by BaSTI.} and 
the errors involved \citep[e.g.,][]{cooke2018}. 

%
%
\section{Conclusions}
\label{sect:conclude}
With the aid of evolutionary and linear stellar pulsation models we computed 
the luminosities of bright RR~Lyrae stars by using basic observed parameters, 
such as period, average magnitudes (Johnson $V$, Gaia $RP$, WISE $W2$), 
iron abundance and interstellar reddening. These luminosities have been 
compared with those derived from the latest release of the Gaia parallaxes. 
By using $\alpha$-enhanced evolutionary models, for the bright part of the 
sample ($62$ stars, with relative parallax errors less than $2$\%), we found 
high level of agreement between the two sets (including both fundamental and 
first overtone pulsators). The $\log L$ values span a range of $\sim 0.5$ 
with an RMS of $0.029$. Solar-scaled models for the same sample yield 
similarly good correlation between the two (essentially independent) luminosity 
sets. However, it seems that the solar-scaled models are brighter by 
$0.012\pm0.004$ in $\log L$. For the remaining $94$ stars with larger parallax 
errors we observe a tendency for increasing Gaia luminosities (relative to 
the theoretical luminosities) both for the $\alpha$-enhanced and solar-scaled 
models. 
     
Based on these results, it seems that stars with accurate parallaxes yield 
well-matching luminosities with the theoretical values {\em without} the 
need for parallax shift. This assertion likely holds whenever the 
`theoretical values' are derived by using:
\begin{itemize}
\item  
$\alpha$-enhanced evolutionary models -- BaSTI \citep{hidalgo2018} or similar; 
\item  
standard radial pulsation models -- e.g., \cite{kovacs2000}; 
\item  
$T_{\rm eff}$ zero point, based on the infrared flux method -- e.g., \cite{gonzalez2009}. 
\end{itemize}  

Our result is essentially in agreement with the conclusion of 
\cite{marconi2021}, based on theoretical period-Wesenheit relations, 
largely relying on nonlinear pulsation models and Gaia DR2 parallaxes. 
In our previous paper on $30$ double-mode RR~Lyrae stars \citep{kovacs2021} 
we landed on a slightly different conclusion, by accepting an overall 
shift of $+0.02$~mas for the EDR3 parallaxes. Because of the considerably 
lower accuracy of data used in that study, this shift may seem to be 
justified, considering the similar trend observed in the present sample. 
However, due to the different method used in our double-mode study 
(because of the lack of metallicity measurements for those faint stars), 
and the size of the sample (together with the large errors involved) 
make us cautious to put strong weight on the conclusion concerning the 
parallax shift from that study.   

Finally, it is important to note that $\sim 10$\% of the stars show curious 
`distance keeping' between the evolutionary and pulsation models (i.e., 
in a strict sense they do not yield theoretical solutions -- with zero 
distance in the $(L,M)$ space, even stretching the errors of the observed 
parameters). We found that an object-dependent increase of the reddening 
(well above the formal error limits) solves the problem. This, however, 
is not very easy to understand, since the reddening maps are believed to 
be based on the total dust density, so any increase in the extinction are 
expected to be related to the star rather than to the interstellar matter.

%
%
\begin{acknowledgements}
%
We thank the referee for the quick and helpful report 
(in particular, suggesting the investigation of the effect 
of Helium abundance). 
This research has made use of the VizieR catalogue access tool, CDS, 
Strasbourg, France (DOI: 10.26093/cds/vizier). 
%
This work has made use of data from the European Space Agency (ESA) 
mission {\it Gaia} (\url{https://www.cosmos.esa.int/gaia}), 
processed by the {\it Gaia} Data Processing and Analysis Consortium 
(DPAC, \url{https://www.cosmos.esa.int/web/gaia/dpac/consortium}). 
Funding for the DPAC has been provided by national institutions, 
in particular the institutions participating in the {\it Gaia} 
Multilateral Agreement.
%
This publication makes use of data products from the Wide-field 
Infrared Survey Explorer, which is a joint project of the University 
of California, Los Angeles, and the Jet Propulsion 
Laboratory/California Institute of Technology, funded by the 
National Aeronautics and Space Administration.
%
This research has made use of the NASA/IPAC Infrared Science Archive, 
which is funded by the National Aeronautics and Space Administration 
and operated by the California Institute of Technology.
%
This research has made use of the International Variable Star Index 
(VSX) database, operated at AAVSO, Cambridge, Massachusetts, USA.
%
Supports from the National Research, Development and Innovation 
Office (grants K~129249 and NN~129075) are acknowledged. 
\end{acknowledgements}

%
%

%

%
\end{document}